\documentclass[aps,prd,11pt,superscriptaddress,nofootinbib]{revtex4-2}
\usepackage[T1]{fontenc}
\usepackage[utf8]{inputenc}
\usepackage[english]{babel}
\usepackage{color}
\usepackage{amsmath}
\usepackage{amssymb}
\usepackage{slashed}
\usepackage{afterpage}
\usepackage{rotating}
\usepackage[toc,page]{appendix}
\usepackage{graphicx}
\usepackage{subcaption}
\usepackage{psfrag}
\usepackage{wrapfig}
\usepackage{array}
\usepackage{tensor}
\usepackage{xfrac}
\usepackage{hyperref}
\usepackage[nameinlink,capitalise]{cleveref}
\usepackage{lineno}
\usepackage{microtype}
\usepackage{needspace}

\newcommand{\So}{\sqrt{3}\Omega}
\newcommand{\Sb}{\sqrt{3}\beta}

\newcommand{\dB}{\dot{\beta}}
\newcommand{\dO}{\dot{\Omega}}
\newcommand{\I}{\indices}

\Crefname{figure}{Fig.}{Figs.}

\crefrangelabelformat{equation}{(#3#1#4--#5\crefstripprefix{#1}#2#6)}

	\makeatletter

\renewcommand{\p@subsection}{}
\makeatother

\begin{document}

\modulolinenumbers[5]
\title{A minisuperspace approach for the Supersymmetric Reissner-Nordstr\"{o}m spacetime}

\author{G. Pollari}
\email{504077@mail.muni.cz}
\affiliation{Department of Theoretical Physics and Astrophysics, Masaryk University, Brno}

\begin{abstract}
\begin{center}
\textbf{Abstract} 
\end{center}
In this work we present a supersymmetric extension of the Reissner-Nordstr\"{o}m spacetime in the framework of quantum cosmology. We achieve this by supersymmetrizing the Wheeler-DeWitt equation for a Kantowski-Sachs cosmological model defined in the interior region between the event and Cauchy horizons. We carry out this analysis following the Dirac and the reduced phase space quantization approaches and compare them. In the quantum regime we describe the novelties introduced by supersymmetry and highlight the differences with the pure bosonic case. At the classical level we analyse some limits that lead to metrics which are different than the usual Reissner-Nordstr\"{o}m form. We show that in some regions of the spacetime, true singularities arise. In addition, we study the geodesics equation of a massive particle moving in the region where supersymmetry contributions are dominant.
\end{abstract}

\maketitle

\section{Introduction}\label{Intro}

The study of black holes has long served as a gateway into the deep structure of gravity and spacetime. Concretely, 
they are widely regarded as prime candidates for probing the fundamental nature of gravity, especially in regimes where quantum
effects become significant. Among the family of classical black hole solutions in general relativity, the Reissner–Nordström (RN) solution 
occupies a central place, describing a static, spherically symmetric charged black hole in four dimensions.
Although astrophysical black holes are likely uncharged, scenarios involving charged black holes have been considered in various
contexts, such as primordial black holes, black holes in strong electromagnetic fields, and as theoretical models to explore black hole thermodynamics 
and their relevance in the context of the holographic duality \citep{Maldacena}.

At the classical level, the causal structure of the RN black hole is characterized by an event horizon and a Cauchy horizon, separating static and dynamical regions of spacetime. 
While traditionally attention is focused on the asymptotically flat region exterior to the outer horizon, it is well known that the RN solution extends beyond the event 
horizon to regions where the coordinates switch their causal character: the radial coordinate becomes timelike, and the temporal coordinate becomes spacelike. 
This interior region can be naturally interpreted as a homogeneous but time-dependent cosmological spacetime.  More generally,  one can exploit the diffeomorphism 
invariance of general relativity and use the change of structure of the spacetime as a technique for generating cosmological solutions from known static (black hole) solutions, 
 by performing the coordinate transformation $t\leftrightarrow r$. This idea has a long history, with early examples including the
Schwarzschild–Kantowski–Sachs cosmology correspondence \citep{kantowski:1966,kuchar:1994}, among others \citep{obregon2004}.

The relation between static and cosmological solutions has profound implications in the context of quantum cosmology, 
where one aims to describe the dynamics of the universe via a wavefunction satisfying the Wheeler–DeWitt
(WDW) equation. In minisuperspace models, the WDW does not distinguish between
exterior (static) and interior (dynamical) regions of a black hole spacetime. This
observation supports the idea that one can define a quantum black hole state via the quantisation of the associated cosmological 
model -a viewpoint explored, for instance, in \citep{Cavaglia:1994yc, Obregonn}. 
 
This idea has also attracted attention from the loop quantum gravity community, where the black hole interior is quantized
using methods developed for cosmological models~\citep{Ashtekar:2005cj}. In string theory, supersymmetry plays a central role in making black holes analytically tractable.
In particular, supersymmetric (BPS) black holes preserve a fraction of supersymmetry, which protects their physical quantities -such as mass and charge- from quantum corrections. 
This protection ensures stability and enables exact microscopic calculations, most notably of black hole entropy via D-brane state counting \citep{Strominger}. These black holes are extremal, namely $M = |Q|$, 
and can be described as classical solutions in supergravity. To incorporate supersymmetry into quantum cosmology, based on the ideas of \citep{Teitelboim:1977fs,Tabensky:1977ic}, 
the pioneering work in~\citep{macias1987quantum} introduced supersymmetric quantum cosmology as the square root of the WDW equation. 
Latter, some other approaches were developed in the literature. A comprehensive review regarding various methods in supersymmetric quantum 
cosmology can be found in \citep{garcia2021topics}. In particular, the work in \citep{Graham:1991av} focuses on constructing supercharges by
identifying a suitable square root of the potential in homogeneous cosmological models, whose anticommutator yields the
Hamiltonian constraint. In this work, we adopt the approach outlined in \citep{Graham:1991av} to construct a supersymmetric
extension of the RN black hole within the quantum cosmology framework, as done in \citep{Lopez-Dominguez:2006rou,Obregon-Zacariasp,Lopez-Dominguez:2011}.
 While retaining a simple framework, this approach provides a tractable and conceptually transparent model where canonical
quantisation and supersymmetry can be studied in a gravitational context. 
To achieve this, we begin by considering the cosmological model obtained from the RN interior via a diffeomorphism that exchanges the static and
dynamical regions. We then study the Hamiltonian formulation of this model and derive the associated WDW equation. 
We do this by considering two quantisation schemes:  Dirac quantisation, where constraints are imposed at the quantum level, and reduced phase
space quantisation, where constraints are solved classically before quantizing. Throughout, we highlight the differences between these two approaches.
We then extend this to the supersymmetric framework by constructing supercharges the square of which leads to the supersymmetric WDW equation, containing additional potential terms associated with the fermionic contributions. Within the minisuperspace approximation, the supersymmetric WDW equation admits closed-form solutions. We analyze the main features of the quantum-level modifications induced by supersymmetry, emphasizing the differences that emerge when compared to the charged Kantowski–Sachs model. In contrast, the corresponding classical equations in this framework do not admit analytical solutions. To make progress, we explore limiting regimes where either the bosonic or fermionic contributions dominate the dynamics. It is worth emphasizing that the causal structure of the spacetime, e.g. the location of the event horizons and the singularity, is lost once the study of the features of geometry within a limited region of validity is restricted. In the bosonic-dominated limit, we recover the standard RN solution whereas in the fermionic-induced dominated regime, we obtain new four classes of solutions. Moreover, in the supersymmetric region, the metrics is purely spacelike and possess singularities depending on the values of the aforementioned effective mass and the electric charge parameters, as revealed in the Kretschmann invariant. Consequently, the study of the nature of the singularities and the implications that could derive from it is indispensable. In this regard, we analyze the geodesic equation of a massive particle radially falling throughout the supersymmetric region. The result demonstrates that those singularities are reached in finite time, marking the end of the spacetime.\vspace{1cm}

\section{WDW equation for the Reissner-Nordstr\"{o}m space-time }\label{SectionRN}
In this section we will construct WDW equations the classical limit of which describes the Reissner-Nordstr\"{o}m space-time. We will do this following different formulations which give rise to generically inequivalent quantum descriptions for the system. We highlight the main differences between them throughout. 
\subsection{The Reissner-Nordstr\"{o}m solution}\label{Classical}
As is well known the solution of Einstein field equations coupled to an electromagnetic field for a spherically symmetric non rotating object is given by the line element
\begin{align}\label{rnr}
ds^2=-\left(1-\frac{2m}{r}+\frac{Q^{2}}{r^{2}}\right)dt^{2}+\left( 1-\frac{2m} {r}+\frac{Q^{2}}{r^2}\right)^{-1}dr^{2}+r^{2}\left(d\theta^{2}+ \sin^{2}\theta d\varphi^{2}\right),  
\end{align}
where $m$ is the black hole (BH) mass and $Q$ its charge the relative values of which determine the causal structure of the spacetime. For $m> \vert Q \vert$ there are two event horizons at $r_{\pm}$. The solution in equation \eqref{rnr} is valid in the regions $0<r<r_{-}$ and $r_{+}< r< \infty$. 
On the other hand, for the region $r_{-}< r< r_{+}$
the signature of the $g_{tt}$ and $g_{rr}$ components is interchanged. If we further perform the coordinate transformation $t\leftrightarrow r$, we find the metric 
\begin{align}\label{rnt}
ds^{2}=-\left(-1 + \frac{2m}{t}-\frac{Q^2
}{t^2}\right)^{-1}dt^{2}+\left(-1 +\frac{2m}{t}-\frac{Q^2}{t^2} \right)dr^{2}+t^{2}\left(d\theta^{2}+\sin^{2}\theta
d\varphi^{2}\right),
\end{align}
which is also solution of the Einstein field equations and can be recognised as the cosmological model associated to the Einstein-Maxwell system and from now on we will refer to this as the charged Kantowski-Sachs cosmology. Moreover, there exists an appropriate way to address the models we have just discussed using the Hamiltonian formulation that we will now review. For the Kantowski-Sachs model we use the Misner parametrization \citep{Misner}
\begin{align}\label{ksmetric}
ds^{2}=-N^{2}dt^{2}+e^{2\sqrt{3}\beta}dr^{2}+e^{-2\sqrt{3}(\beta + \Omega)}\left(d\theta^{2}+\sin^{2}\theta
d\varphi^{2}\right). 
\end{align}
where $N$  denotes the lapse function and $\beta$, $\Omega$ time-dependent functions.
 The Einstein-Maxwell action, choosing units where $2\kappa=1$, reads
 $$S= \int_{\mathcal{M}} d^4x \: \sqrt{-g} \left( R - F_{\mu \nu}F^{\mu \nu}\right). $$
Since the metric depends only on time, the vector potential $A_\mu$ must also depend solely on time to satisfy the Einstein field equations (EFE). Consequently, spatial derivatives are dropped, and we choose an ansatz where only the radial component of $A_\mu$  is non-zero. Using this, the ADM action becomes
\begin{equation} \label{Action}
    S = \int_{\Sigma} \int_{\mathbf{R}} d^3x \, dt \: \sqrt{q} \left( \dfrac{6}{N} (\dot{\beta}^2 - \dot{\Omega}^2) +  \dfrac{2}{N} q^{rr} \dot{A}_r \dot{A}_r + N R\I{^{(3)}}\right),
\end{equation}
 where $\sqrt{q}$ is the determinant of the induced metric on spatial slices, and $R\I{^{(3)}}$ is the 3-dimensional Ricci scalar.
 The corresponding momenta conjugate to the fields are
 \begin{align}\label{0} 
 &&P_\beta = \dfrac{12}{N} \sqrt{q}\dot{\beta}, &&P_\Omega = -\dfrac{12}{N} \sqrt{q} \dot{\Omega}, &&P_A = \dfrac{4}{N} \sqrt{q} q^{rr}\dot{A}_r, &&\Pi = 0, 
 \end{align}
with the canonical pairs satisfying the Poisson brackets
 \begin{align}\label{1} 
 &&\{P_\beta,\beta\} = 1, && \{P_\Omega,\Omega\}=1, &&\{P_A, A_r\} = 1, &&\{\Pi,N\} = 1. 
 \end{align}
The Hamiltonian constraint turns out to be 
\begin{equation}\label{2}
      P_\beta^2 - P_\Omega^2 + 3e^{2\sqrt{3}\beta}P_A^2 - 48e^{-2\sqrt{3}\Omega} = 0. 
 \end{equation}
The vector potential is constrained by its equation of motion, which we obtain by varying \eqref{Action} with respect to $A_r$,
\begin{equation}\label{3}
    \dfrac{d}{dt}\left( \dfrac{1}{N} e^{-3\sqrt{3}\beta -2\sqrt{3}\Omega} \dot{A_r}\right) = 0, 
\end{equation}
which is solved by
\begin{equation} \label{4}
    F_{tr} = \dot{A_r} = Q N e^{3\sqrt{3}\beta +2\sqrt{3}\Omega} \;\; ,
\end{equation}
and using \eqref{0} we obtain
\begin{equation} \label{5}
    P_A=4Q,
\end{equation}
where $Q$ is a constant that can be interpreted as the charge. Furthermore, by choosing $t^2=e^{-2\sqrt{3}(\beta+\Omega)}$, it is straightforward to derive the condition $N^2 = e^{-2\sqrt{3}\beta}$. Using this value for the lapse, together with  
 \eqref{0} and \eqref{5} into \eqref{2}, we obtain the equation
\begin{equation}\label{10}
    e^{-2\sqrt{3}\Omega}\left(1+2\sqrt{3}\dot{\Omega} t \right) + Q^2 -t^2=0,
\end{equation}
the solution of which is
\begin{equation}\label{11}
    e^{-2\sqrt{3}\Omega} = -t^2 +2 m t -Q^2,
\end{equation}
where $m$ is a constant associated to the mass of the charged object. It follows that 
\begin{equation}\label{12}
    e^{2\sqrt{3}\beta} = - 1+\dfrac{2m}{t}-\dfrac{Q^2}{t^2}, 
\end{equation}
and
\begin{equation}\label{13}
    N^2 = -\left(- 1+\dfrac{2m}{t}-\dfrac{Q^2}{t^2}\right)^{-1}.
\end{equation}
Using \eqref{ksmetric}, the metric reads
\begin{equation}\label{14}
    ds^2 = -\left(-1 + \dfrac{2m}{t} - \dfrac{Q^2}{t^2}\right)^{-1} \; \:dt^2 + \left( -1 + \dfrac{2m}{t} - \dfrac{Q^2}{t^2} \right) \:dr^2 + t^2 (d\theta^2 +\sin^2\theta d\varphi^2),
\end{equation}
which is nothing but the solution in \eqref{rnt} and after interchanging $t\leftrightarrow r$ and moving to either $r_{+}< r< \infty$ or $0<r<r_{-}$ region the one in \eqref{rnr}. 

\subsection{The WDW equation}\label{wdw}
In the quantisation procedure, two principal approaches can be taken: Dirac quantisation (DQ) and Reduced phase space quantisation (RPSQ) \citep{DQ-RQ} \footnote{see also \citep{Thiemann-RPS} for an extensive and rigorous approach applied to gravity.}. In DQ, the canonical variables are promoted to quantum operators and the classical Poisson brackets replaced by commutators. Constraints in this approach are maintained and solved at the quantum level. In RPSQ, by contrast, constraints or equations of motion, are partially or entirely resolved at the classical level. Certain canonical variables are expressed as functions of others, reducing the system’s degrees of freedom before quantisation. Only the remaining independent degrees of freedom are promoted to quantum operators. The difference in how the phase space is constrained and quantized in DQ and RPSQ generally leads to different results, reflecting the distinct choices made in handling constraints at each stage. We will discuss both approaches in what follows. 
\\ \noindent
In the Dirac quantisation picture the Hamiltonian constraint \eqref{2} and \eqref{4} become
\begin{subequations}\label{15}
    \begin{align}
        &\left(\dfrac{\partial^2}{\partial \Omega^2} - \dfrac{\partial^2}{\partial \beta^2} - 3e^{2\sqrt{3}\beta} \dfrac{\partial^2}{\partial A^2} - 48e^{-2\sqrt{3}\Omega} \right) \Psi(\Omega,\beta,A) \approx 0, \label{15a}\\
        & -i\dfrac{\partial}{\partial A} \Psi(\Omega,\beta,A) = \nu \Psi(\Omega,\beta,A), \label{15b}
    \end{align}
\end{subequations}
where $\nu$ is a real constant. The solution of the system \eqref{15} is given by
\begin{equation}\label{16}
    \Psi_{\lambda,\nu}(\Omega,\beta,A) = e^{i\nu A} \:K_{i\lambda} \left(4 e^{-\sqrt{3}\Omega}\right ) K_{i \lambda}\left(\nu e^{\sqrt{3}\beta}\right), 
\end{equation}
where $K$ is the modified Bessel function of the second kind and $\nu,\lambda$ are separation constants \citep{Cavaglia1}. \\ \noindent
On the other hand, in the Reduced phase space approach we use equation \eqref{5}, which was derived from the classical equation of motion for the system. We therefore substitute \eqref{5} directly into \eqref{2} to obtain the WDW equation 
\begin{equation}\label{17}
        \left(\dfrac{\partial^2}{\partial \Omega^2} - \dfrac{\partial^2}{\partial \beta^2} + 48 \left(Q^2 e^{2\sqrt{3}\beta} - e^{-2\sqrt{3}\Omega}\right) \right) \Psi(\Omega,\beta) = 0,
\end{equation}
that gives the solution
\begin{equation}\label{18}
    \Psi_\lambda(\Omega,\beta) = K_{i\lambda} \left(4 e^{-\sqrt{3}\Omega}\right ) K_{i \lambda}\left(4 Q e^{\sqrt{3}\beta}\right).
\end{equation}
The wavefunctions \eqref{16} and \eqref{18} differ by a phase factor with frequency $\nu$ that reflects the $U(1)$ invariance of the electromagnetic action. In the RPSQ, the gauge vector is entirely determined by the geometry, hence the absence of the phase factor and its associated frequency $\nu$ is related to the charge of the electromagnetic field solving $\eqref{15b}$ at the classical level $P_A=\nu =4Q$. The wavefunction \eqref{16} describes a Kantowski–Sachs “multiverse’’ in which both the charge $\nu$ and the gravitational oscillation frequency $\lambda \in \mathbb{R}_+$ are summed over. The universe sits in a quantum superposition of all possible values of the electric momentum and all Bessel‐mode excitations. On the other hand, in the wavefunction \eqref{18}, the classical electric charge $Q$ is fixed classically before quantisation and only the gravitational modes $\lambda$ remain dynamical. The aim of this work is to supersymmetrise the aforementioned WDW equations in the spirit of \citep{Graham:1991av} (see also \citep{DEATH199344}), that we shall interpret as the supersymmetric WDW equations for the RN black hole following the ideas in \citep{Obregon-Zacariasp, Lopez-Dominguez:2011}.

Before to continue, it is useful to study the properties of the wavefunction \eqref{18} and construct wave packets out of it. In the following we choose a Gaussian wave packet of the form
\begin{equation}\label{18.1}
    \Psi = \int_0^\infty d\mu_1 \ldots \int_0^\infty d\mu_n  \: e^{-\frac{(\mu_1-\bar{\mu}_1)^2}{2\sigma_1^2}} \ldots e^{-\frac{(\mu_n-\bar{\mu}_n)^2}{2\sigma_n^2}}\Psi_{\mu_1\ldots\mu_n}, 
\end{equation}
where $\mu_1, \ldots,\mu_n$ are the frequencies, $\bar{\mu}_1, \ldots,\bar{\mu}_n$ the expectation values and $\sigma^2_1, \ldots,\sigma^2_n$ the variances.
\begin{figure}[t!]
\centering
\begin{subfigure}[b]{0.5\textwidth}
\centering
    \includegraphics[width=8cm]{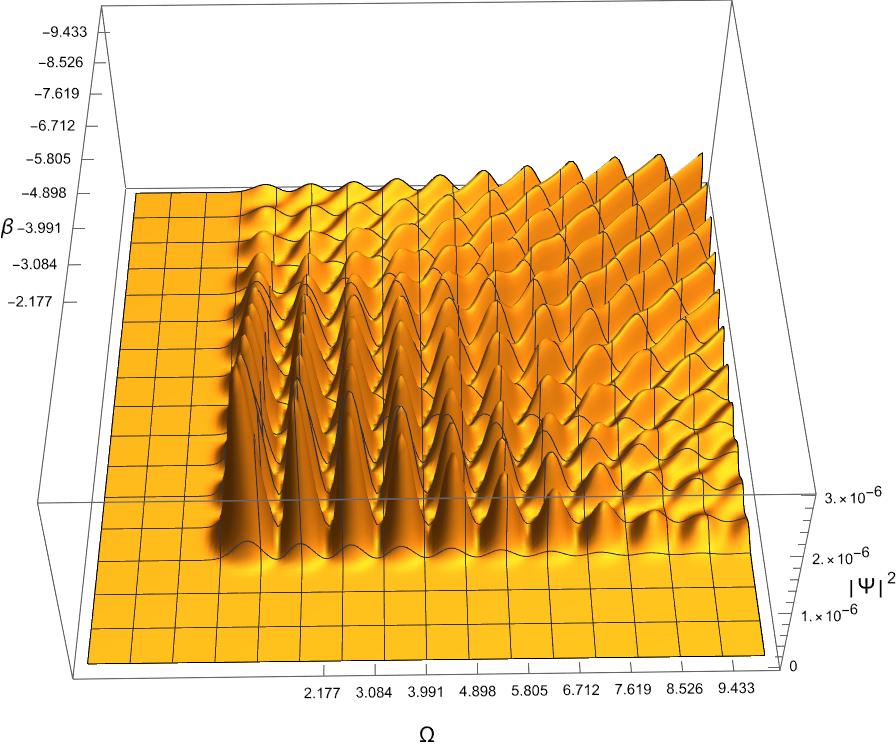}
    \caption{Variation of $|\Psi|^2$ with respect to $\Omega$ and $\beta$ for $\bar{\lambda}=2$, $\sigma^2=0.01$ and $Q=1$.}
    \label{fig:Fig.1a}
    \end{subfigure}
    \hfill
    \begin{subfigure}[b]{0.45\textwidth}
    \centering
    \includegraphics[width=7.5cm]{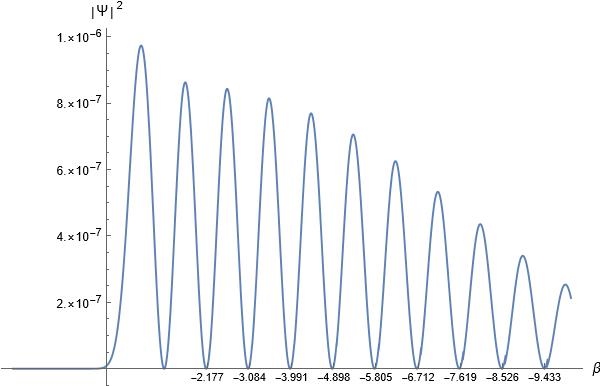}
    \caption{Variation of $|\Psi|^2$ with respect to $\beta$  at $\Omega=4.44$ for $\bar{\lambda}=2$, $\sigma^2=0.01$, $Q=1$.}
    \label{fig:Fig.1b}
    \end{subfigure}
    \label{fig:Fig.1}
\end{figure}

\begin{figure}[t!]
\centering
    \includegraphics[width=8cm]{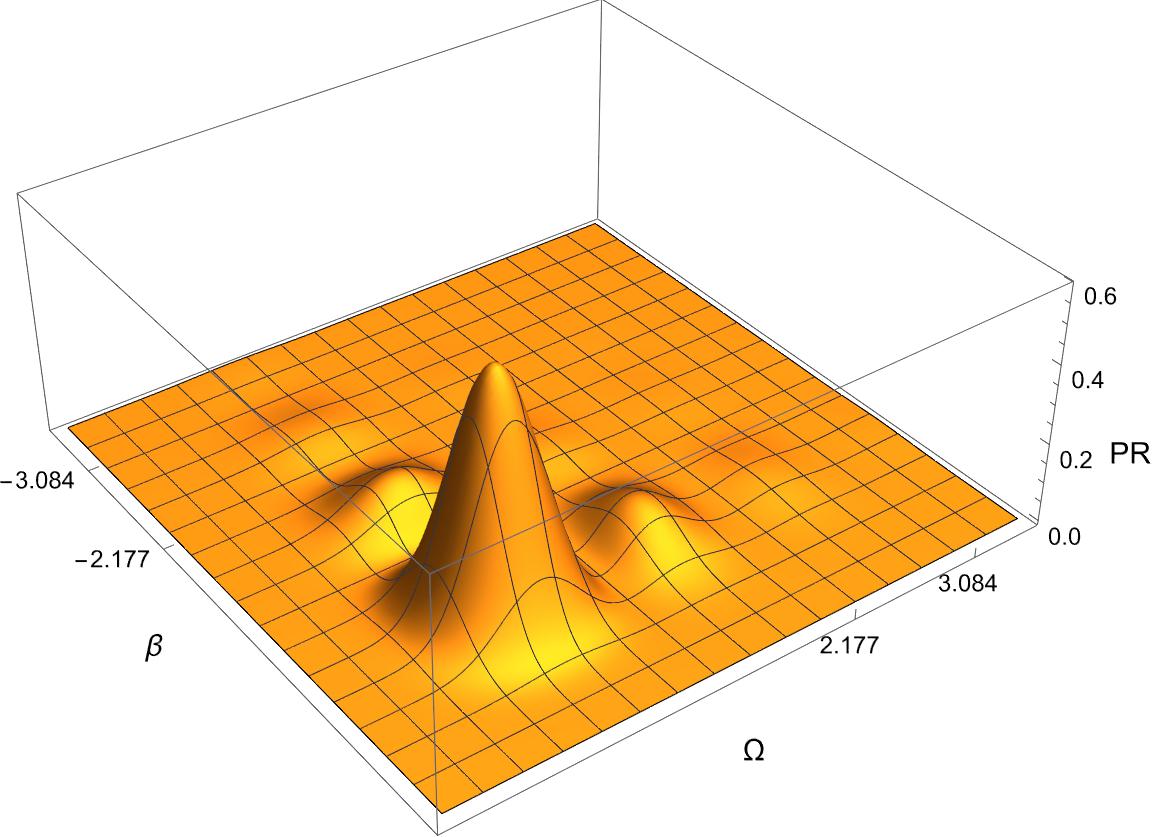}
    \caption{Probability density $PR$ for $\lambda=2$ and $Q=1$.}
    \label{fig:Fig.2}
\end{figure}

The Bessel function $K_{i\lambda} \left(4 e^{-\sqrt{3}\Omega}\right)$ decays rapidly as $\Omega \rightarrow -\infty$ while it oscillates an infinite number of times for $\Omega \gg 1$. By taking the asymptotic expansion, we find
$$K_{i\lambda} \left(4 e^{-\sqrt{3}\Omega}\right) \approx 2^{-1-i\lambda} \left(4 e^{-\sqrt{3}\Omega}\right)^{-i\lambda} \left( \left(4 e^{-\sqrt{3}\Omega}\right)^{2 i\lambda} \Gamma(-i \lambda) + 2^{2 i \lambda} \Gamma( i \lambda)\right), \hspace{1cm} \Omega\rightarrow +\infty.$$
The values of $\Omega$ at which it vanishes are found to be
$$ \Omega_{k_1} = - \dfrac{1}{\sqrt{3}} \ln \left( \dfrac{1}{2} e^{-\frac{\pi k_1}{\lambda}} \left(\dfrac{\Gamma (-i \lambda)}{\Gamma(i\lambda)}\right)^{\frac{i}{2\lambda}}\right)  \hspace{2.5cm} k_1\in \mathbb{N},$$
from which the oscillation frequency is $f=\pi/(\sqrt{3}\lambda)$. For our analysis, we choose $\lambda=2$, which yields a frequency of $f \approx 0.907$. The same approach applies for $\beta$, with the only difference being the presence of the charge $Q$. The zeros of $\beta$ are given by
$$\beta_{k_2} = \dfrac{1}{\sqrt{3}} \ln \left( \dfrac{1}{2Q} e^{-\frac{\pi k_2}{\lambda}} \left(\dfrac{\Gamma (-i \lambda)}{\Gamma(i\lambda)}\right)^{\frac{i}{2\lambda}}\right) \hspace{2.5cm} k_2\in \mathbb{N}.$$
Hence, the charge shifts the zeros along the $\beta$ axis but does not alter the frequency. Each node (zero point) of the wavefunction occurs at the intersection point between the line joining two consecutive local maxima along one axis and the line intersecting the midpoints between two consecutive local maxima along the second axis, as illustrated in \autoref{fig:Fig.1a}. The second graph \autoref{fig:Fig.1b} depicts the zeroes of the wavefunction along the $\beta$ axis at $\Omega=4.44$. The nodes have a clear interpretation. In the WKB approximation the oscillatory part corresponds to a positive momentum, while the decay part corresponds to the classical prohibited zone. Take standard ansatz $\Psi \approx \text{exp}(i(\int P_\Omega d \Omega + \delta)$ where the phase is $\varphi(\Omega) = \int P_\Omega d\Omega + \delta$. For $\Omega \gg 1$, using the asymptotic expansion derived above, it follows that $P_\Omega = \sqrt{3} \lambda$. The zeroes of the wavefunction occur at the points $\Omega_k$ whenever $\varphi(\Omega_k) = P_{\Omega} \Omega_k + \delta = k \pi$. Inverting for $\Omega$ we find $\Omega_k= (k \pi -\delta)/\sqrt{3}\lambda$. Thus, the difference between two consecutive nodes $\Delta\Omega= \Omega_{k+1}- \Omega_{k}= \pi/\sqrt{3}\lambda$, which is exactly the frequency we found above, is the amount needed for the phase to increase by $\pi$. Notice, however, that since the Wheeler–DeWitt potential is unbounded from below, it possesses only one finite turning point; therefore no quantisation condition arises, $\lambda$ remains continuous, and the fixed $\pi$-phase advance between successive nodes represents a scattering phase shift rather than a bound-state quantum number. This analysis shows that there exist infinitely many periodic values of $\Omega$ and $\beta$ where the wavefunction vanishes
$$ \Psi\left((\Omega_{k_1} + \Omega_{k_1 +1})/2,\beta_{k_2}\right)= \Psi\left((\Omega_{k_1},(\beta_{k_2}+\beta_{k_2+1})/2\right)=0, $$
indicating that a universe is not possible for those specific configurations.
Although the wavefunction is not normalizable in the classical sense, we would like to choose a well-defined scalar product. To this purpose, according to \citep{Pal-Banerjee}, we make the following change of variables $X= 4e^{-\sqrt{3}\Omega}$, $Y=4Q e^{\sqrt{3}\beta}$ and introduce the function $\phi_\lambda(X) = K_{i\lambda}(X)$. Equation \eqref{18} then transforms to
$$\Psi_\lambda(X,Y) = \phi_\lambda(X) \phi_\lambda(Y).$$
We define the scalar product as follows:
$$ \langle \Psi_{\lambda_1} | \Psi_{\lambda_2} \rangle \equiv \int_0^\infty \int_0^\infty dX dY \: X Y \Psi^*_{\lambda_1}(X,Y) \Psi_{\lambda_2}(X,Y) =  \dfrac{\pi^4}{4}\left(\dfrac{\lambda_1^2 - \lambda_2^2}{\cosh(\pi \lambda_1) -\cosh( \pi \lambda_2) }\right)^2, $$
for any $\lambda_1 \neq 0$,$\lambda_2 \neq 0$. The induced norm is given by
$$ || \Psi_{\lambda} ||^2=  \int_0^\infty dX  \: X |\Psi_{\lambda}(X)|^2 \int_0^\infty dY \: Y |\Psi_{\lambda}(Y)|^2 < \infty.$$
Hence, $\Psi_\lambda \in L_2^X[0,\infty) \oplus  L_2^Y[0,\infty)$, the direct sum of weighted $L_2$ spaces with weight $X$ and $Y$, respectively. The Gaussian wave packet of the probability density $PR(X,Y) = X Y |\Psi_{\lambda}(X)|^2 |\Psi_{\lambda}(Y)|^2$ in terms of $\beta,\Omega$ for $\lambda=2$ is illustrated in \autoref{fig:Fig.2}.
It indicates that the expected values predominantly lie around $\Omega \in [0,3]$, $\beta \in [-3,0]$ whereas the region outside such an interval has been washed out compared to \autoref{fig:Fig.1a}. Although the wavefunction is not normalizable under the standard scalar product, the probability density determines the relative likelihood of different regions in the configuration space. In our case, it quantifies how much more likely it is for the universe to exist in one configuration compared to another.

\section{A supersymmetric WDW equation for the Reissner-Nordstr\"{o}m spacetime.}\label{SectionSSDQ}

In this section we will generalize the WDW equations \eqref{15a} and \eqref{17} to its supersymmetric regime. In order to achieve this, we will follow the proposal presented in \citep{Graham:1991av,Obregon-Zacariasp}, which -as discussed in the introduction- relies on finding a "square root" of the potential of the standard (bosonic) model. To be more concrete, for the cosmological model described by a Hamiltonian written in the compact form
\begin{equation} \label{WDW-gen}
2\hat H_0= G^{\mu\nu}\, P_\mu P_\nu + U(q),
\end{equation}
where $q$ are the generalized coordinates, $G^{\mu\nu}$ the minisuperspace metric and $U(q)$ the potential, we seek a function $\phi$ which is a solution of the Hamilton-Jacobi equation
    \begin{gather}
G^{\mu \nu}\frac{\partial \phi}{\partial \phi^{\mu}}\frac{\partial \phi}
{\partial q^{\nu}} = U(q)
\label{H-J}
\end{gather}
that defines the supersymmetric Hamiltonian constraint as follows
\begin{gather}
H=\frac{1}{2}\left\{Q,\overline{Q}\right\} =
H_0+\frac{1}{2}\frac{\partial^2 \phi}{\partial q^\mu\partial
q^\nu}\left[\bar{\theta}^\mu,\theta^\nu\right],\label{superhamiltonian}
\end{gather}
where Q are the supercharges for which in a differential representation take the form
\begin{align}
Q=\theta^{\mu}\left(\Pi_{\mu}+ i\frac{\partial \phi}{\partial
q^{\mu}}\right),~~
\overline{Q}=\overline{\theta}^{\mu}\left(\Pi_{\mu}
-i\frac{\partial \phi}{\partial q^{\mu}}\right),\label{supc}
\end{align}
where the Grassmannian variables $\theta$ satisfy the algebra
\begin{align}
\left\{\bar{\theta}^{\mu},\bar{\theta}^{\nu}\right\}=0,~~~~
\left\{\theta^{\mu},\theta^{\nu}\right\}=0,~~~~
\left\{\bar{\theta}^{\mu},\theta^{\nu}\right\}=G^{\mu\nu},
\label{algebragrass}
\end{align}
where in our model $G^{\mu \nu} = \text{diag}(-1,1,3e^{2\sqrt{3}\beta})$. In order to satisfy this constraint algebra, we will chose a matrix representation for the aforementioned variables. A convenient choice in terms of the Pauli matrices $\sigma_i, \; i=1,2,3$ and $\sigma_\pm=\frac{1}{2}\left(\sigma_1 \pm i\sigma_2\right)$ is given by
\[\begin{array}{ll}
\hat{\theta}^{\Omega}=\sigma_{-}\otimes I\otimes I & \hspace{1cm}\hat{\bar{\theta}}^{\Omega}=-\sigma_{+}\otimes I \otimes I\nonumber\\
\hat{\theta}^{\beta}=\sigma_{3}\otimes\sigma_{-} \otimes I & \hspace{1cm} \hat{\bar{\theta}}^{\beta}=\sigma_{3}\otimes\sigma_{+} \otimes I\\
\hat{\theta}^{A}=\sqrt{3}e^{\sqrt{3} \beta} \left(\sigma_{3} \otimes \sigma_{3} \otimes \sigma_{-}\right) & \hspace{1cm} \hat{\bar{\theta}}^{A}= \sqrt{3}e^{\sqrt{3} \beta}\left(\sigma_{3} \otimes \sigma_{3} \otimes \sigma_{+} \right) .\label{sigmas}
\end{array}\]

As we did for the pure bosonic case,  we will now construct the supersymmetric WDW equation following the two different approaches outlined in Section \ref{wdw}.

\subsection{Supersymmetric Dirac quantisation}\label{SubSectionSSDQ}
\begin{figure}[t!]
        \centering
        \includegraphics[width=0.43\linewidth]{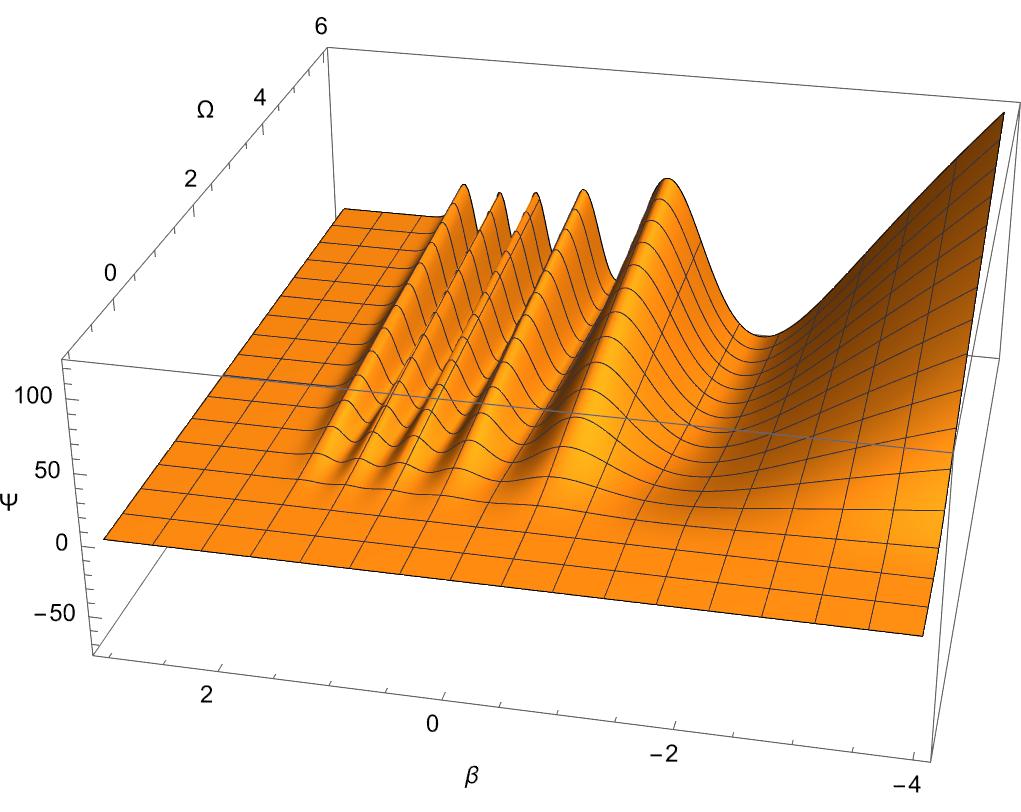}
        \caption{Graph of the SUSY wavefunction for $A=const.$, $\lambda=0$, $\omega=10$, $\nu=1$. The turning points are at $\Omega_t \approx 1.2$, $\beta_t \approx 1.73$.}
        \label{fig:Whittaker}
    \end{figure}

Using the minisuperspace metric $G^{\mu \nu} = \text{diag}(-1,1,3e^{2\sqrt{3}\beta})$ w.r.t. the coordinates $(\Omega,\beta,A)$, equation \eqref{H-J} gives 
\begin{equation}\label{SuperspaceHam}
    -\left(\dfrac{\partial \phi}{\partial \Omega} \right)^2+ \left(\dfrac{\partial \phi}{\partial \beta}\right)^2 + 3e^{2\sqrt{3}\beta} \left(\dfrac{\partial \phi}{\partial A}\right)^2 + 48e^{-2\sqrt{3}\Omega} =0.
\end{equation}
A solution is: 
\begin{equation}\label{solphi}
    \phi(\Omega,\beta,A) = -4 e^{- \sqrt{3} \Omega} + 4 \omega e^{ \sqrt{3} \beta} + 4 i \omega A,
\end{equation}
where $\omega$ is a constant. Note that for $\omega=0$ the solution \eqref{solphi} reduces to the one found in \citep{Obregon-Zacariasp}, i.e. to the chargeless case. This means that such a constant is somehow connected to the electromagnetic charge $\nu$ because turning off $A$, $\omega$ disappears. However, in the full quantum theory we cannot establish the function which relates them. The annihilation of $\Psi$ by the minisuperspace Hamiltonian \eqref{superhamiltonian} then takes the form
\begin{gather} \label{42}
\left(
\begin{array}{cccccccc}
H_0 + f_1 &0&0&0 &0&0&0 &0\\
0 &H_0 + f_1 &0&0&0 &0&0&0\\
0& 0& H_0 + f_2&0&0 &0&0&0\\
0& 0& 0& H_0 + f_2&0&0 &0&0\\
0&0&0&0 &H_0 - f_2&0 &0 &0\\
0&0&0&0 &0 &H_0 - f_2&0 &0\\
0&0&0 &0&0&0 &H_0 - f_1 &0\\
0&0&0 &0&0&0&0 &H_0 - f_1
\end{array}
\right)
\left(
\begin{array}{c}
\Psi_1\\ \Psi_2\\ \Psi_3\\ \Psi_4 \\ \Psi_5\\ \Psi_6\\ \Psi_7\\ \Psi_8
\end{array}
\right)=0 
\end{gather}
where $$f_1 (\Omega,\beta) = 6 \left( e^{-\sqrt{3}\Omega} + \omega e^{\sqrt{3}\beta} \right) \hspace{1cm} f_2 (\Omega,\beta) = 6 \left(- e^{-\sqrt{3}\Omega} + \omega e^{\sqrt{3}\beta} \right)$$
which is independent of $A$. Now, we have a set of eight equations, yet half of them are redundant, therefore only four of them describe a different state of the universe. While it is analytically possible to solve the quantum equations presented above, transitioning to the classical level to obtain the corresponding metric proves to be unfeasible. Consequently, it becomes necessary to consider asymptotic limits, that will be discussed in the next section. In order to understand the contribution due to supersymmetry on the wave function, we solve the WDW equations \eqref{42}
\begin{equation}\label{42.1}
    \left(\dfrac{\partial^2}{\partial \Omega^2} - \dfrac{\partial^2}{\partial \beta^2} - 3e^{2\sqrt{3}\beta} \dfrac{\partial^2}{\partial A^2} - 48e^{-2\sqrt{3}\Omega} +6 (\pm e^{-\sqrt{3}\Omega} \pm \omega e^{\sqrt{3}\beta}) \right) \Psi(\Omega,\beta,A) \approx 0.
\end{equation}
Using the method of separation of variables $\Psi(\Omega,\beta,A) = f(\Omega) g(\beta) h(A)$ we obtain the system
\begin{subequations}
   \begin{align}
     &  \left(\frac{\partial^2}{\partial\Omega^2} - 48 e^{-2\sqrt{3}\Omega} \pm 6 e^{-2\sqrt{3}\Omega} + 3\lambda^2\right) f(\Omega)=0 \label{TPO}\\ 
     & \left(\frac{\partial^2}{\partial\beta^2} - 3 \nu^2 e^{2\sqrt{3}\beta} \pm 6 \omega e^{\sqrt{3}\beta} + 3\lambda^2 \right) g(\beta)=0  \label{TPB} \\
     &\left(\frac{\partial^2}{\partial A^2} + \nu^2 \right) h(A)=0
\end{align} 
\end{subequations}
whose solution is given by
\begin{equation} \label{43}
   \Psi^{\scalebox{0.7}{SUSY}}_{\lambda, \nu,\omega}(\Omega, \beta, A) = e^{i \nu A} e^{\frac{\sqrt{3}}{2} (\Omega - \beta)} W_{\pm \frac{1}{4}, i \lambda}(8 e^{-\sqrt{3}\Omega}) W_{\pm \frac{\omega}{\nu}, i \lambda}(2 \nu e^{\sqrt{3}\beta}), 
    \end{equation}
where $W$ is the Whittaker function. The positive indices $1/4$ and $\omega/\nu$ are specifically associated with the function $f_1$, the negative indices $-1/4$ and $-\omega/\nu$ with the function $-f_1$. The appearance of $\omega$ in \eqref{solphi} coupled to $A$ 
and in \eqref{43} as an index together with $\nu$, suggests it could be interpreted as a frequency, potentially interconnected with the electromagnetic charge. We shall come to this below. By consistency \eqref{43} reduces to \eqref{16} when the original bosonic potential dominates over the fermionic one. This occurs in the asymptotic regime $\Omega \rightarrow -\infty, \: \beta \rightarrow +\infty$ where the supersymmetric contributions become negligible, and the standard (bosonic) wavefunction is recovered. Specifically, we have:
\begin{align}\label{44}
   & e^{\frac{\sqrt{3}}{2} \Omega} W_{\pm \frac{1}{4},i \lambda}(8 e^{-\sqrt{3}\Omega}) \approx e^{-4 e^{-\sqrt{3}\Omega}} \approx K_{i\lambda}(4 e^{-\sqrt{3}\Omega}), &\Omega\rightarrow -\infty \\
 &e^{-\frac{\sqrt{3}}{2} \beta} W_{\pm \frac{\omega}{\nu},i \lambda}(2 \nu e^{\sqrt{3}\beta}) \approx e^{\nu e^{\sqrt{3}\beta}} \approx K_{i\lambda}(\nu e^{\sqrt{3}\beta}), &\beta\rightarrow +\infty
\end{align}
 thus matching the form of the bosonic wavefunction \eqref{16}.\\ \noindent
As done in Section \ref{SectionRN}, the zeroes of the function in \eqref{44}, calculated via asymptotic expansion, are determined by the points
 \begin{equation}\label{44.a}
  \Omega_{k_1} = -\dfrac{1}{\sqrt{3}} \ln{ \left(\dfrac{1}{8}  e^{ -\frac{\pi k_1}{\lambda}} \left(\dfrac{\Gamma(1/4-i\lambda)\Gamma(2i\lambda)}{\Gamma(1/4+i\lambda)\Gamma(-2i\lambda)}\right)^{-\frac{i}{2\lambda}}\right)}, \hspace{3cm} k_1 \in \mathbb{R},
  \end{equation}
 \begin{equation}\label{44.b}
 \beta_{k_2} = \dfrac{1}{\sqrt{3}} \ln{ \left(\dfrac{1}{2\nu}  e^{ -\frac{\pi k_2}{\lambda}} \left(\dfrac{\Gamma(1/2-i\lambda -\omega/\nu)\Gamma(2i\lambda)}{\Gamma(1/2+i\lambda-\omega/\nu)\Gamma(-2i\lambda)}\right)^{-\frac{i}{2\lambda}}\right)}, \hspace{2.5cm} k_2 \in \mathbb{R},
 \end{equation}
 where the frequency matches that of the bosonic case. \\ \noindent
 More compelling it is how supersymmetry affects the location of the turning points of the universe. A turning point is the value of the configuration variable at which the classical momentum vanishes, i.e. where the wavefunction switches from oscillatory (classically allowed) to exponential (classically forbidden) behaviour. In the bosonic charged Kantowski-Sachs universe, the Bessel function $K_{i \lambda} (x)$ oscillates for $x<\lambda$, decays for $x>\lambda$ and $x=\lambda$ is the turning point \citep{CAMPBELL199149}. In particular, $\lambda=0$ carries no oscillation, therefore no classically allowed region. In the supersymmetric universes, the wavefunctions are the Whittaker functions \eqref{43}. The turning points are the points where the effective potentials in \crefrange{TPO}{TPB} vanish. Setting $x=8 e^{-\sqrt{3}\Omega}$ and $y= 2\nu e^{\sqrt{3}\beta}$, we obtain
 \begin{subequations}
 \begin{align}
     &x_t^\pm = \frac{\pm 1 + \sqrt{1+18 \lambda^2}}{2} \label{TPx} \\
     & y_t^\pm = \pm \frac{2 \omega}{\nu} + 2\sqrt{\frac{\omega^2}{\nu^2}+ \lambda^2}.\label{TPy}
     \end{align}
 \end{subequations}
where the $\pm$ is associated to the signs of the effective potentials. It can be easily shown that the turning points $x_t^+$ and $y_t^+$ are pushed further to the right with respect to the bosonic case, whereas $x_-$ is pushed to left and $y_t^-$ to the left if $\lambda<3 \omega/4\nu$ or to the right if $\lambda>3 \omega/4\nu$. Moreover, a compelling feature is that for $\lambda=0$, $x_t^+$ and $y_t^+$ no longer vanish. This means that the wavefunction 
\begin{equation} \label{45}
   \Psi^{\scalebox{0.7}{SUSY}}_{\lambda, \nu,\omega}(\Omega, \beta, A) = e^{i \nu A} e^{\frac{\sqrt{3}}{2} (\Omega - \beta)} W_{\frac{1}{4}, i \lambda}(8 e^{-\sqrt{3}\Omega}) W_{\frac{\omega}{\nu}, i \lambda}(2 \nu e^{\sqrt{3}\beta}) 
    \end{equation}
    starts oscillating for $\Omega>\sqrt{3}\ln 2\equiv \Omega_t$, $\beta<\ln(2\omega/\nu^2)/\sqrt{3}=\beta_t$. The domain $(\Omega,\beta) \in (\Omega_t,\infty) \times(-\infty,\beta_t)$ is the set of values of the gravitational variables for which classical dynamics is permitted. 
    Consequently, the supersymmetric extension allow the universe described by the wavefunction \eqref{45} to have a classical dynamics in the 0-mode, which is conversely forbidden in the bosonic framework. The wavefunction \eqref{45} is plotted in Fig.\ref{fig:Whittaker}.
\subsection{Supersymmetric Reduced phase space quantisation}\label{SubsectionSSRD}
In the RPSQ we solve for the electric momentum before quantizing. Using \eqref{0}, the key difference is the independence of $A$ and therefore the minisuperspace metric is $G^{\mu \nu} = \text{diag}(-1,1)$. Consequently, \eqref{SuperspaceHam} becomes 
\begin{equation}\label{SuperspaceHam2}
    -\left(\dfrac{\partial \phi}{\partial \Omega} \right)^2+ \left(\dfrac{\partial \phi}{\partial \beta}\right)^2 + 48 (e^{-2\sqrt{3}\Omega} - Q^2 e^{2\sqrt{3}\beta})=0.
\end{equation}
which yields 
\begin{equation}\label{solphi2}
    \phi(\Omega,\beta,A) = -4 e^{- \sqrt{3} \Omega} + 4 Q e^{ \sqrt{3} \beta}.
\end{equation}
The constant $\omega$ in \eqref{solphi} is entirely fixed in terms of the electric charge as $\omega=Q$. The wavefunction arising from this reduced model is obtained by identifying $\nu =4 \omega=4Q$ in  \eqref{43}:
\begin{equation}\label{44.2}
     \Psi^{\scalebox{0.7}{SUSY}}_{\lambda}(\Omega, \beta)=e^{\frac{\sqrt{3}}{2} (\Omega - \beta)} W_{\pm \frac{1}{4}, i \lambda}(8 e^{-\sqrt{3}\Omega}) W_{\pm \frac{1}{4}, i \lambda}(8 Q e^{\sqrt{3}\beta}).
     \end{equation}
    The points at which \eqref{44.2} vanishes can be readily obtained from \eqref{44.a} and \eqref{44.b} by making the aforementioned substitutions. \\ \noindent
    Moreover, the modification induced by supersymmetry does not cure the the normalization issue of the wavefunction as the potential is still unbounded from below. A well-defined scalar product is defined in the same way as constructed in Section \ref{wdw}. Changing variables according to  $X= 8 e^{-\sqrt{3}\Omega}$ and $Y=8Q e^{\sqrt{3}\beta}$, the inner product reads
$$ \langle \Psi^{\scalebox{0.7}{SUSY}}_{\lambda_1} | \Psi^{\scalebox{0.7}{SUSY}}_{\lambda_2} \rangle \equiv \int_0^\infty \int_0^\infty dX dY \: X Y \Psi^{\scalebox{0.7}{SUSY}^*}_{\lambda_1}(X,Y) \Psi^{\scalebox{0.7}{SUSY}}_{\lambda_2}(X,Y).$$
The induced norm is found to be
$$ || \Psi^{\scalebox{0.7}{SUSY}}_{\lambda} ||^2=  \int_0^\infty dX  \: X |\Psi^{\scalebox{0.7}{SUSY}}_{\lambda}(X)|^2 \int_0^\infty dY \: Y |\Psi^{\scalebox{0.7}{SUSY}}_{\lambda}(Y)|^2 < \infty.$$
Therefore, the wavefunction is again $\Psi^{\scalebox{0.7}{SUSY}}_{\lambda,\omega} \in L_2^X[0,\infty) \oplus  L_2^Y[0,\infty)$. In the next section we shall discuss the classical limit, which brings a significant shift from the charged Kantowski-Sachs universe.
\subsection{\textbf{The classical limit}}\label{SectionCL}
\begin{figure}[h]
\centering
\begin{subfigure}[t]{0.48\textwidth}
\centering
    \includegraphics[width=8.2cm]{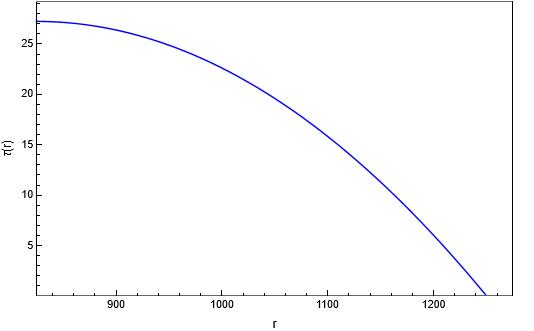}
    \caption{\scalebox{0.85}{Geodesic of a massive particle for $V_{(+,+)}$ from $r_0$ to $r_{1}$.}}
    \label{fig:Fig.3a}
    \end{subfigure}
    \hfill
    \begin{subfigure}[t]{0.48\textwidth}
    \centering
    \includegraphics[width=8.2cm]{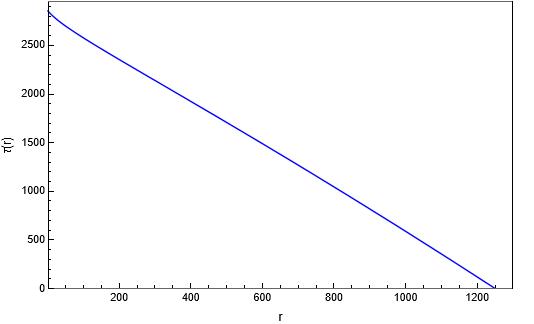}
    \caption{\scalebox{0.83}{Geodesic of a massive particle for $V_{(-,-)}$ from $r_0$ to \scalebox{0.89}{$r=0$}.}}
    \label{fig:Fig.3b}
    \end{subfigure}

    \vspace{0.5cm}

\begin{subfigure}[t]{0.48\textwidth}
\centering
    \includegraphics[width=8.2cm]{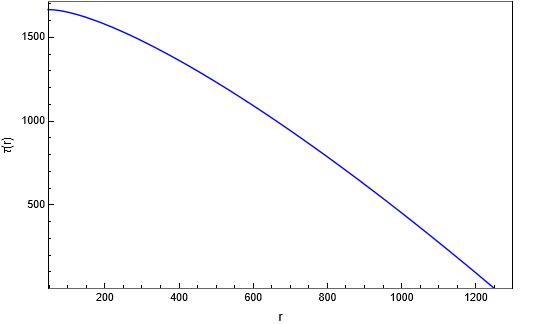}
    \caption{\scalebox{0.85}{Geodesic of a massive particle for $V_{(-,+)}$ from $r_0$ to $r_{-}$.}}
    \label{fig:Fig.3c}
    \end{subfigure}
    \hfill
    \begin{subfigure}[t]{0.48\textwidth}
    \centering
    \includegraphics[width=8.2cm]{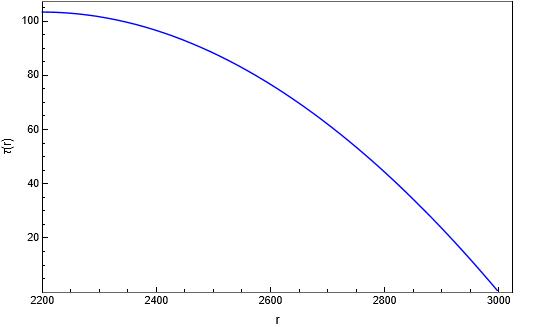}
    \caption{\scalebox{0.85}{Geodesic of a massive particle for $V_{(+,-)}$ from $r_0$ to $r_{+}$.}}
    \label{fig:Fig.3d}
    \end{subfigure}
    \label{fig:Fig.3}
\end{figure}

In the previous section, we compared the DQ procedure with the RPSQ one. Although these two routes differ only minimally, they do not yield exactly the same quantum theory. Since any valid quantisation must reproduce a single, well-defined classical theory in the limit $\hbar\to0$, we adopt the reduced phase space approach as it is partially on-shell. This implies that the parameters $\nu, \omega$ featuring in the DQ must be fully determined by the electric charge $Q$ as derived earlier. \\ \noindent
The starting point is the quantum constraint
\begin{equation}
\hat H\,\Psi(\beta,\Omega)
=\Bigl[-\hbar^2\,G^{\mu\nu}\,\partial_\mu\partial_\nu + U(\beta,\Omega)\Bigr]\Psi(\beta,\Omega)
=0.
\label{WDW-operator}
\end{equation}
Let $\Psi \sim e^{\frac{i}{\hbar}S}$ be the standard WKB ansatz, where $S$ is the Hamilton's principal function. Assuming $S(\Omega,\beta)= S_1(\Omega)+S_2(\beta)$, inserting $\Psi$ into \eqref{WDW-operator} and retaining only the $\mathcal{O}(\hbar^0)$ term it gives the Hamilton-Jacobi equation
$$G^{\mu\nu}\,\partial_\mu S\,\partial_\nu S + U(\beta,\Omega) = 0.$$
Identifying the momenta as $P_\mu=\partial_\mu S$ we obtain 
\begin{equation}\label{46}
    P_\beta^2 - P_\Omega^2 +U_{Cl} + U_{SUSY} =0,
\end{equation}
where the classical potential is $U_{Cl}= 48( Q^2 e^{2\sqrt{3}\beta} -  e^{-2\sqrt{3}\Omega})$ and the SUSY potential is $U_{SUSY}= 6 (\pm Q e^{\sqrt{3}\beta} \pm e^{-\sqrt{3}\Omega})$.
 Eq. \eqref{46} differs from the classical constraint \eqref{2} by the presence of the additional $U_{SUSY}$ originating from the incorporation of supersymmetry. The SUSY potential grows/ decays more slowly than its bosonic counterpart, thus its imprint on asymptotic limits of the geometrical variables is tiny. However, for some parameter configurations, it could be the dominant  potential within a finite region of the spacetime, yielding a disparate metric with respect to the classical one. We were unable to find analytical solution to this equation, and as such, in order to understand the influence of the supersymmetric generalization we will study the equation by taking two limiting cases. In Table \ref{Table1} we have displayed the limits we will consider in what follows.
Since $e^{-2\sqrt{3}(\Omega+\beta)}=t^2$, the metric M.1 corresponds to a finite region wherein the Reissner–Nordström geometry dominates and the two additional terms stemming from supersymmetry are suppressed while M.2 describes the geometry in the domain where the supersymmetric potential fully governs the dynamics (for details on the derivation, see Appendix \ref{APPENDIXA}). In this last case, the $\pm$ signs correspond to the different signs appearing in the supersymmetric potential $U_{SUSY}$ and $c_2$ is a positive constant of integration. It is worth emphasizing that, similarly to the bosonic solution, $c_2$ could be interpreted as a mass parameter associated to the mass encoded in the spacetime. In other words, as for the charge $\omega$ which is locked-in to the electromagnetic charge, the parameter $c_2$ should be a function of the ADM mass $m$. However, it must be noted that the supersymmetric region is confined to a finite region. Once causality is reversed by interchanging $t$ and $r$, it becomes evident that this region is spacelike. Therefore, the function $c_2(m)$ expressing $c_2$ with $m$ cannot be firmly established. Since the pure SUSY local spacetime resides in the dynamical region of the full spacetime, it can be interpreted as a perturbation to the classic Reissner-Nordstr\"{o}m geometry. 
\begin{table}[b]
    \begin{tabular}{|m{4em}|m{10cm} |m{0.9cm}|}
    \hline
    \scalebox{1}{\( U_{Cl}\)}& \scalebox{1.1}{ \( ds^2=-\left(-1 + \frac{2m}{t} -\frac{Q^2}{t^2}\right)^{-1}dt^{2}+\left(-1 +\frac{2m}{t} -\frac{Q^2}{t^2}\right)dr^{2} \)} &  \scalebox{1}{M.1}\\
    \hline
    \scalebox{1}{\(U_{SUSY} \)}& \scalebox{1.1}{\( ds^2 = -8t \:dt^2 +\left( \pm 1 -\frac{2 c_2}{\sqrt{t}} \pm \frac{Q}{t}\right)^2 \)}$dr^2$  &  \scalebox{1}{M.2}\\
    \hline
    \end{tabular}
    \caption{Metric from the asymptotic behaviors of SUSY extension \eqref{46}. In the two blocks: M.1) classical charged Kantowski-Sachs, M.2) supersymmetric region.}
    \label{Table1}
    \end{table}
    It shifts the location of the event horizons by a tiny amount, yet, in the narrow volume supersymmetry prevails, it generates new features. In fact, although the metric in M.2 is regular, the Kretschmann invariant \eqref{A.2} reveals that it possesses true singularities at
\begin{gather}
  t_{\scalebox{0.7}{(+,+)}} = \left(c_2 \pm \sqrt{c_2^2 - Q}\right)^2 \tag{S.1}\\
  t_{\scalebox{0.7}{(+,\scalebox{1.7}{-})}} = \left(c_2 + \sqrt{c_2^2 + Q}\right)^2  \tag{S.2} \\
  t_{\scalebox{0.7}{(\scalebox{1.7}{-},+)}} = \left(c_2 - \sqrt{c_2^2 + Q}\right)^2  \tag{S.3}
\end{gather}
while $t_{\scalebox{0.84}{(\scalebox{1.25}{-},\scalebox{1.25}{-})}} \notin \mathbb{R}$. The indices \scalebox{0.84}{$(\pm,\pm)$} correspond to the four signs in $U_{SUSY}$. The solution S.1 admits two singularities if $c_2 > |Q|$, one singularity if $c_2 = |Q|$ or none if $c_2< |Q|$. In contrast, both S.2 and S.3 always exhibit one singularity each. In order to better understand the implications of the SUSY metric we examine the geodesics of a massive particle undergoing radial motion after interchanging causality by $t \leftrightarrow r$ (see Appendix \ref{APPENDIXB} for its derivation). In our analysis we choose $c_2 = 20$, $Q=324$ and $E=100$. The graphs of the solution of \eqref{B.4} for $\tau(r)$ are illustrated in \autoref{fig:Fig.3a}{-d}.
The singularity for $V_{\scalebox{0.7}{(+,+)}}$ lies at $r_1=824.7$ \footnote{The inner singularity at $r_2=127.3$ is not a region of the domain since the validity of the spacetime terminates at $r_1$.}, whereas $V_{\scalebox{0.7}{(\scalebox{1.5}{-},\scalebox{1.5}{-})}}$ does not show any singularity. On the other hand, $V_{\scalebox{0.7}{(+,\scalebox{1.5}{-})}}$ and $V_{\scalebox{0.7}{(\scalebox{1.5}{-},+)}}$ exhibit a single singularity located at $r_+=2200$ and $r_-=47.7$, respectively. Consider \autoref{fig:Fig.3a} as an example. Such a massive particle starts at $r_0=1250$ in proper time $\tau_0=0$ with an arbitrary velocity. It accelerates till it reaches the outer singularity at $r_1$ in a finite time. In the given coordinate chart, all singularities occur at finite radius and are topologically $S^2$. These are surface-like singularities, and the spacetime manifold terminates at them — no extension beyond these spheres is possible within classical general relativity. This stands in contrast to the curvature singularity at $r=0$ in the Reissner–Nordström black hole, which is a single point.

\section{Conclusions and outlooks}\label{Conclusions}

In this paper we have studied a minisuperspace quantisation for the interior region of the Reissner–Nordström interior and its supersymmetric extension. We briefly summarized the well-know classical diffeomorphism $t\leftrightarrow r$ in the Reissner-Nordstr\"{o}m metric that enables a cosmological study in the Hamiltonian picture. Starting from the Kantowski–Sachs form of the charged black‐hole metric, we compared two quantisation paths: \textit{Dirac quantisation}, in which the $U(1)$ electromagnetic constraint remains unfixed at the quantum level, yielding a WDW wavefunction that superposes over all electric momentum eigenvalues and the \textit{Reduced phase‐space quantisation}, where the charge is fixed classically before quantizing. We constructed Gaussian wave-packets selecting a specific Bessel-mode $\lambda$ and defined a finite scalar product to visualize the density probability of the quantum states. Subsequently, by applying the Graham–D’Hoker approach, we constructed non‐Hermitian supercharges which produce a supersymmetric WDW operator containing additional fermionic‐induced potentials. The supersymmetric-extended WDW equation is analytically solved by the superposition of oscillating Whittaker functions with mode $\lambda$. In particular, in the DQ picture, it emerged an arbitrary constant $\omega$ that in the RPSQ corresponds to the electric charge. In the asymptotic limit $\Omega\to -\infty,\;\beta\to+\infty$, where the bosonic potential prevails over the SUSY potential, these reduce to the bosonic Bessel‐function wavefunctions, thus ensuring agreement with the non‐SUSY models in their common regime of validity. Furthermore, we showed how the addition of SUSY shifts the turning points of the original charged Kantowski-Sachs universe, enlarging or shrinking the range of the classically allowed/forbidden universe. In particular, there exists one SUSY potential which switches the zero-mode of the universe wavefunction to be classically allowed in a specific region, marking a net distinction with the non-SUSY model. Taking the standard WKB ansatz we extracted the effective classical Hamilton–Jacobi equation which, in addition to the usual RN potential $U_{Cl}$ features a milder “SUSY correction” $U_{SUSY}$. The differential equation does not admit a closed-form solution, hence we solved the resulting equations in two limits: bosonic‐dominated regime and SUSY‐dominated regime. The former corresponds to regions which include the asymptotically flat region and extends slightly beyond the location of the classical outer horizon, while the latter emerges in a narrow domain which lies entirely within the so-called dynamical region, where the roles of time and space reverse, i.e. the radial coordinate becomes time-like. The SUSY potentials, corresponding to the four possible combinations of the signs, exhibit novel features. Three potentials give rise to true singularities $(S.1),(S.2),(S.3)$ whereas the other potential does not. We confirmed such results by studying the geodesic equation of a massive particle in radial falling within the SUSY domain. The singularities are topologically 2-spheres, constitute the end of the classical regime of General Relativity and as such the spacetime cannot be extended further. \\ \noindent
A possible extension of our work is the study of junction conditions, in particular performing a detailed Israel junction \citep{Israel} matching between the pure RN region and the SUSY-dominated interior, computing the induced surface stress–energy tensor and exploring whether any physically reasonable thin shell can connect the two patches smoothly. Moreover, an analysis about the implications of supersymmetry in the quantum framework shift of “turn‑on” radius for quantum behaviour. Investigating how this displacement affects the onset of quantum cosmological bounces \citep{Bojo1} \citep{Chiou} or tunnelling probabilities \citep{VILENKIN198225} \citep{Fanaras_2022} would shed light on the physical consequences of the supersymmetric extension.
\section*{Acknowledgments}
This work was supported by the grant MUNI/A/1503/2024 of Masaryk University and in part by the grant GA23-06498S of Czech Science Foundation.\\ 
I would like to thank Salomon Gabriel Zacarias Nicasio for advancing the ideas presented here and engaging fruitful discussions.\\ 
Finally, good luck to Salomon Gabriel Zacarias Nicasio—aka \textit{El Pera}—on his valiant and intrepid mission to cycle from Brno to Cape Town. All the \textit{Perpetue} of the Theoretical Physics and Astrophysics department are with you!

\begin{appendices}
\numberwithin{equation}{section}
\section{Classical limit in the SUSY regime}\label{APPENDIXA}
The Hamiltonian constraint in SUSY domination reads:
$$ P_\beta^2 - P_\Omega^2 +6 (\pm Q e^{\sqrt{3}\beta} \pm e^{-\sqrt{3}\Omega})=0$$
which stems from the Lagrangian
 $$S = \int_{\Sigma} \int_{\mathbf{R}} d^3x \, dt \: \left( \dfrac{6}{N} e^{-\Sb -2\So} (\dot{\beta}^2 - \dot{\Omega}^2) \mp \dfrac{N}{4}Q e^{2\sqrt{3}(\beta+\Omega)} \mp \dfrac{N}{4} e^{\sqrt{3}(\beta+\Omega)}\right).$$
The corresponding EoM are
\begin{subequations}\label{SYSTEM1}
\begin{align}
    & \dfrac{6}{N} e^{-\Sb -2 \So} (\dot{\beta}^2- \dot{\Omega}^2) \pm \dfrac{N}{4}Q e^{2\sqrt{3}(\beta+\Omega)} \pm \dfrac{N}{4} e^{\sqrt{3}(\beta+\Omega)}=0 \label{N1}\\
    & \pm \dfrac{N}{16} Q e^{2\sqrt{3}(\beta+\Omega)} +\sqrt{3} \dfrac{d}{dt}\left( \dfrac{1}{N}  e^{-\Sb -2 \So}\:\dot{\beta} \right)=0 \label{b1}\\
    & \pm \dfrac{N}{16}  e^{\sqrt{3}(\beta+\Omega)}+\sqrt{3} \dfrac{d}{dt}\left( \dfrac{1}{N}  e^{-\Sb -2 \So}\: \dot{\Omega} \right)=0. \label{O1}
\end{align}
\end{subequations}
Summing \eqref{b1} and \eqref{O1} we obtain
$$ \pm \dfrac{N}{16} Q e^{2\sqrt{3}(\beta+\Omega)} + \pm \dfrac{N}{16} e^{\sqrt{3}(\beta+\Omega)} +\sqrt{3} \dfrac{d}{dt}\left( \dfrac{1}{N}  e^{-\Sb -2 \So}\:(\dB+ \dO) \right)=0, $$
and using \eqref{N1} it becomes
$$2\sqrt{3} \dfrac{d}{dt} \ln\left( \dfrac{1}{N}  e^{-\Sb -2 \So}\: (\dB +\dO) \right) -3 \left( \dB - \dO\right)=0,$$
 leading to
 \begin{equation}\label{NSUSY}
    N = -\dfrac{1}{\sqrt{3}k_1} e^{-\frac{\sqrt{3}}{2} (\beta+ \Omega)} \dfrac{d}{dt} e^{-\sqrt{3} (\beta +\Omega)}.
 \end{equation}
 In order to simplify the system above we can choose $e^{-2\sqrt{3}(\beta+\Omega)} =t^2$. This is possible because the lapse function is a Lagrangian multiplier leaving a free DoF in \eqref{SYSTEM1} that can be 
arbitrary chosen. Thus, \eqref{NSUSY} reduces to
 \begin{equation}\label{NSUSY2}
    N = - \dfrac{\sqrt{t}}{\sqrt{3}k_1},
 \end{equation}
 and substituting \eqref{NSUSY2} into \eqref{b1} we obtain the following differential equation:
 $$ \dfrac{d}{dt} \left( t^{3/2} \dfrac{d}{dt} e^{\Sb}\right) \pm \dfrac{Q}{48 k_1^2}t^{-3/2}=0, $$
 which yields 
 \begin{equation}\label{betaSUSY}
    e^{2 \Sb} =  \left( \pm \dfrac{Q}{24 k_1^2 t} -\dfrac{2 c_2}{\sqrt{t}} \pm c_3\right)^2.
 \end{equation}
 Making use of \eqref{NSUSY} and \eqref{betaSUSY}, eq.\eqref{O1} gives the following ODE 
    $$\dfrac{d}{dt} \left( \sqrt{t} \dfrac{d}{dt}e^{- \So}\right) \mp \dfrac{1}{48 k_1^2 \sqrt{t}}=0$$
whose solution is
     \begin{equation}\label{OmegaSUSY}
     e^{-2 \So} = \left( \pm \dfrac{t}{24 k_1^2} - 2 k_2 \sqrt{t} \pm k_3\right)^2 .
\end{equation}
Some constants can be fixed by using \eqref{betaSUSY} and \eqref{OmegaSUSY} in order to satisfy $e^{-2\sqrt{3}(\beta+\Omega)} =t^2$. It is easy to show that $c_2=k_2$, $k_3 = Q c_3$ and $24 k_1^2 c_3 =1$. Choosing $k_1= - 1/\sqrt{24}$ we obtain the metric
\begin{equation}\label{A.1}
    ds^2 = -8t \:dt^2 + \left( \pm 1 -\dfrac{2 c_2}{\sqrt{t}} \pm \dfrac{Q}{t}\right)^2 \: dr^2 .
\end{equation}
Since such a metric holds only in a region far from $t=0$, it is everywhere regular within the validity of the asymptotic regime. However, the Kretschmann invariant given by 
\begin{equation}\label{A.2} K = \dfrac{39 + 64t(1+4t) + \dfrac{t(44 c_2^2 \mp 76 c_2 \sqrt{t} +33t)}{(\pm t -2c_2 \sqrt{t} \pm Q)^2} + \dfrac{2(38 c_2 \sqrt{t} \mp 33t)}{(\pm t -2c_2 \sqrt{t} \pm Q)}}{64t^6}
\end{equation}
reveals some potential singularities.
\section{The geodesic equation}\label{APPENDIXB}
Upon swapping $t \leftrightarrow r$, the Lagrangian of a massive free particle in the supersymmetric metric 
$$ ds^2 = \left( \pm 1 -\dfrac{2 c_2}{\sqrt{r}} \pm \dfrac{Q}{r}\right)^2 \: dt^2 -8r \:dr^2 + r^2 \: d\Omega^2$$reads
\begin{equation}\label{B.1}
    2 \mathcal{L} = \left( \pm 1 -\dfrac{2 c_2}{\sqrt{r}} \pm \dfrac{Q}{r}\right)^2 \: \dot{t}^2 -8r \: \dot{r}^2 + r^2 \left( \dot{\theta}^2 + \sin^2 \theta \: \dot{\phi}^2 \right)
\end{equation}
where the dot indicates the derivative with respect to the proper time $\dot{x^\mu} \equiv dx^\mu /d\tau$. We shall study the radial free fall, namely $\phi=const$ and $\theta=\pi/2$.
\begin{enumerate}
  \renewcommand{\labelenumi}{\textbf{\arabic{enumi})}}
  \item \textbf{Equation for t:} \\
\begin{center}$\dfrac{\partial\mathcal{L}}{\partial t}=0 $\hspace{2cm} $\dfrac{\partial\mathcal{L}}{\partial \dot{t}}= 2 \dot{t} \left( \pm 1 -\dfrac{2 c_2}{\sqrt{r}} \pm \dfrac{Q}{r}\right)^2$,
\end{center}
which gives 
\begin{equation}\label{B.2}
    \dot{t} = \dfrac{E}{ \left( \pm 1 -\dfrac{2 c_2}{\sqrt{r}} \pm \dfrac{Q}{r}\right)^2},
\end{equation}
where $E$ is the conserved energy per unit mass of the test particle within the SUSY region associated to the time-translation Killing vector $\partial_t$.\\
\item \textbf{Equation for r:} \\
It is convenient to derive it from $\mathcal{L}= g_{\mu \nu} \dot{x^\mu} \dot{x^\nu} = -1$ which yields
$$ \left( \pm 1 -\dfrac{2 c_2}{\sqrt{r}} \pm \dfrac{Q}{r}\right)^2 \dot{t}^2 - 8 r \dot{r}^2 =-1.$$
Making use of \eqref{B.2} we obtain
\begin{equation}\label{B.3}
    \dfrac{dr}{d\tau}= -\dfrac{1}{\sqrt{8 r}} \sqrt{\dfrac{E^2}{\left( \pm 1 -\dfrac{2 c_2}{\sqrt{r}} \pm \dfrac{Q}{r}\right)^2}+1}
\end{equation}
which, once solved for $\tau(r)$, describes the time measured by the particle as it travels radially inward. For brevity, we recast \eqref{B.3} as
\begin{equation}\label{B.4}
    \dfrac{dr}{d\tau}= -V_{(\pm,\pm)}(r)
\end{equation}
where the $\pm$ signs correspond to the signs of the first and last term in the brackets on the R.H.S. of \eqref{B.3}. For example
$$V_{(+,+)}(r)= \dfrac{1}{\sqrt{8 r}} \sqrt{\dfrac{E^2}{\left( 1 -\dfrac{2 c_2}{\sqrt{r}} + \dfrac{Q}{r}\right)^2}+1}$$
and the other three cases follow accordingly.
\end{enumerate}

\end{appendices}

\bibliographystyle{apsrev4-2}
\end{document}